\pdfoutput=1
\RequirePackage{color}
\RequirePackage{ifpdf}

\documentclass{JINST}

\usepackage{color}

\title{A steerable UV laser system for the calibration of liquid argon time projection chambers}

\author{A. Ereditato$^a$, I. Kreslo$^a$, M. L\"uthi$^a$, C. Rudolf von Rohr$^a$, M. Schenk$^a$,
T. Strauss$^a$\thanks{Corresponding author.}, M. Weber$^a$ ~and M. Zeller$^a$\\
\llap{$^a$}Laboratory for High Energy Physics (LHEP), Albert Einstein Center for Fundamental Physics (AEC), University of Bern\\ Sidlerstrasse 5,3012 Bern, Switzerland,\\
E-mail: \email{thomas.strauss@lhep.unibe.ch}}

\abstract{A number of liquid argon time projection chambers (LAr TPC's) are being build or are proposed for neutrino experiments on long- and short baseline beams. For these detectors a distortion in the drift field due to geometrical or physics reasons can affect the reconstruction of the events. Depending on the TPC geometry and electric drift field intensity this distortion could be of the same magnitude as the drift field itself. Recently, we presented a method to calibrate the drift field and correct for these possible distortions. While straight cosmic ray muon tracks could be used for calibration, multiple coulomb scattering and momentum uncertainties allow only a limited resolution. A UV laser instead can create straight ionization tracks in liquid argon, and allows one to map the drift field along different paths in the TPC inner volume. Here we present a UV laser feed-through design with a steerable UV mirror immersed in liquid argon that can point the laser beam at many locations through the TPC. The straight ionization paths are sensitive to drift field distortions, a fit of these distortion to the linear optical path allows to extract the drift field, by using these laser tracks along the whole TPC volume one can obtain a 3D drift field map. The UV laser feed-through assembly is a prototype of the system that will be used for the MicroBooNE experiment at the Fermi National Accelerator Laboratory (FNAL).}

\DeclareGraphicsExtensions{.pdf,.png,.jpg}

\keywords{Neutrino detectors,Time projection chambers,Detector alignment and calibration methods (lasers, sources, particle-beams), Detector design and construction technologies and materials}

\usepackage{lineno}

\begin{document}
\section{Introduction}
\begin{figure}[b]
\begin{center}
\includegraphics[height=8.5cm]{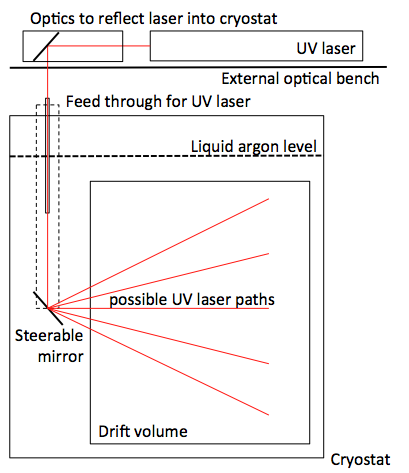}
\caption{Concept of the UV laser calibration system. A UV laser provides light beams; an optical system directs these into a liquid argon cryostat, where a steerable cold mirror allows to direct the laser beams at different paths into the liquid argon volume.}\label{fig:concept}
\end{center}
\end{figure}

The liquid argon time projection chamber (LAr TPC), originally proposed in \cite{rubbia77}, is a prime technology for the detection of neutrino interactions in short- and long baseline beam experiments \cite{microboone,lar1,lbno,lbne}. Large volume detectors with uniform readout and different sizes and shapes can be built with this technique. Recent developments, such as signal pre-amplifiers located in the liquid at 87\,K \cite{bnl1,bnl2}, allow one to design cost effective devices suitable to detect neutrino interactions with sub-millimeter space accuracy on the track reconstruction. On the way towards very massive detectors, the MicroBooNE \cite{microboone} experiment in the Booster Neutrino Beam (BNB) at Fermi National Accelerator Laboratory (FNAL) is going to be commissioned soon for investigating the low energy excess of electron events previously measured by the MiniBooNE \cite{miniboone1,miniboone2,miniboone3,miniboone4} and LSND \cite{lsnd} experiment. Its LAr TPC detector will be able to distinguish the energy (charge) deposition of photon conversions from single-electron tracks, reducing one of the major physics backgrounds coming from $\nu_\mu$ charge current interactions with $\pi_0$ emission \cite{lowenergyexcess1,lowenergyexcess2,lowenergyexcess3,lowenergyexcess4}. 

\begin{figure}[t]
\begin{center}
\includegraphics[height=7cm]{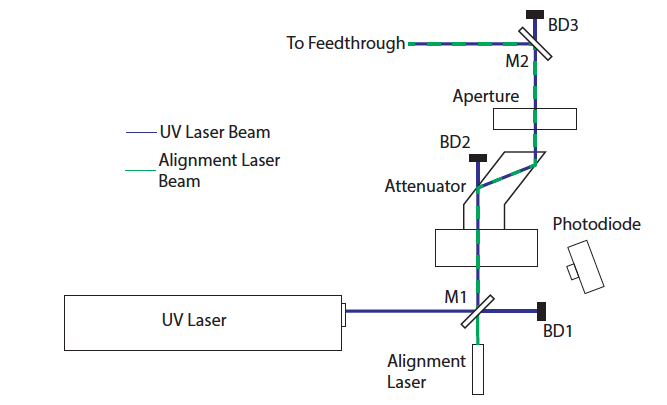}
\caption{A schematic drawing of the external optical bench. An alignment laser (visible light) can be introduced at the first dichroic mirror M1. The attenuator adjusts the beam intensity and the beam diameter is controlled by an aperture. A motorized mirror M2 deflects the beam into the feed-through. Beam dumps (BD) are installed to absorb the non-reflected laser light.}
\label{fig:opticalbench}
\end{center}
\end{figure}

\begin{figure}[h]
\begin{center}
\includegraphics[height=8cm]{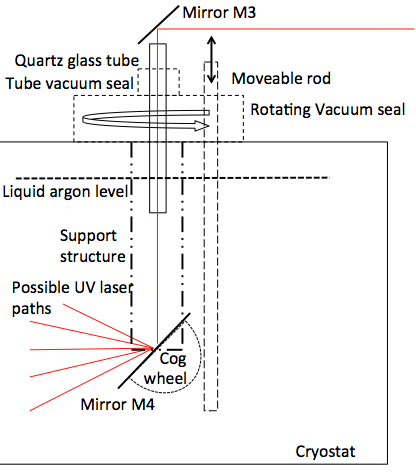}
\caption{A schematic drawing of the feed-through. The UV laser path enters from the right and reflects onto a dichroic mirror M3 into the quartz glass pipe. The bottom of this pipe is well below the liquid level to avoid defocusing of the beam. The whole feed-through assembly can be rotated around the center axis. A moveable rod extending to a cog wheels at the lower mirror M4 allows another degree of freedom. A support structure is fixing the mirror to the top flange of the feed-through.}
\label{fig:feedthrough}
\end{center}
\centering	
\includegraphics[width=0.8\linewidth]{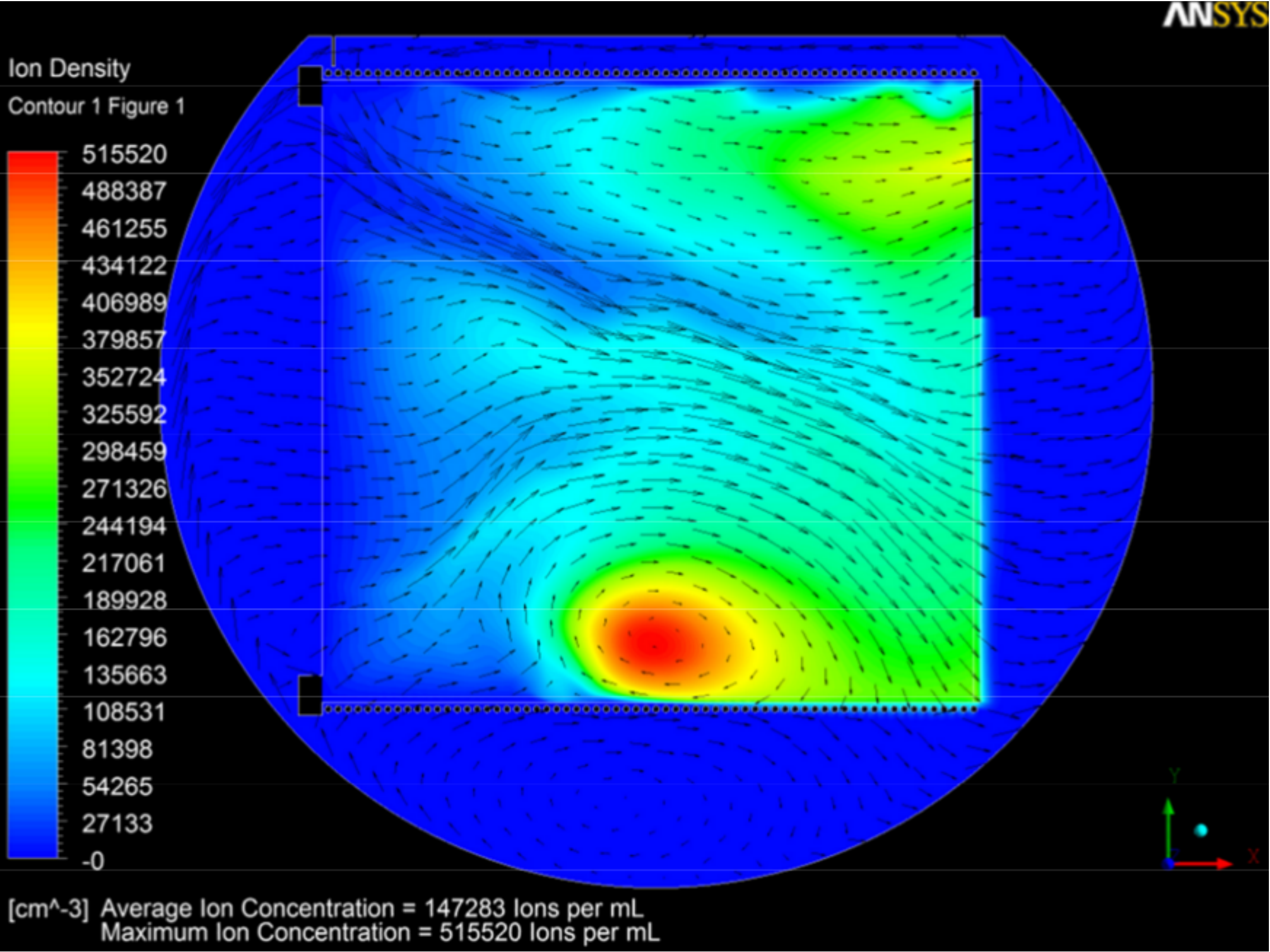}
\caption{Distribution of the positive space charge in presence of argon circulation.}
\label{Voirin}
\end{figure}

Since the MicroBooNE detector is placed at ground level, space charge effects induced by the constant rate of cosmic ray muons are an issue. Space charge is created by the abundance of positively charged ions in the TPC, since the ion drift velocity is five orders of magnitude slower than for electrons. Previous studies \cite{Voirin1} show that the ion concentration can be very non-uniform, depending on the convective flow of liquid argon. Shown in Figure \ref{Voirin} is the ion distribution inside the TPC volume for the MicroBooNE experiment. Local ion accumulation can lead to local electric fields with the same magnitude as the nominal drift field. Thus, the drifting electrons produced by ionizing particles might experience different drift speeds along their path to the readout, distorting the shape of the charge distribution and hence reducing the track reconstruction capabilities of the detector. Therefore, a precise mapping of these local drift field is required to correct for the distortions and maintaining the design performance of the device. 

At the LHEP in Bern a novel method for the calibration of LAr TPC with long drift distance was successfully developed \cite{igor} by using a high intensity UV laser beam to ionize liquid argon \cite{uvlaser1,uvlaser2}. Straight UV induced laser tracks were used to measure the argon purity and the true electric field along the  5\,m long drift of the ARGONTUBE LAr TPC \cite{argontube,phdzeller}. For these measurements, a UV feed-through was used allowing for one single path into the LAr TPC volume. A UV laser track creates a straight ionizing path into the TPC. Local dis-uniformities of the electric field will distort the straight track during the drift towards the sensing readout. As shown in \cite{igor} the electric field can be extracted from the data obtained by UV laser tracks when fitting the track with a straight line. For the MicroBooNE experiment with a length of 10\,m and the used UV laser with a divergence of 0.5mrad the pointing accuracy needs to be better than 5\,mm. In order to perform a full mapping of the drift field, ionizing tracks spanning the whole TPC volume are required. This can be done by employing a mirror immersed in the liquid argon volume to steer the external laser beam. In Section \ref{sec2} we explain the concept and the design of a steerable UV calibration device, while in Section \ref{sec3} we show first performance results obtained with a test TPC device operated at LHEP Bern. 

\section{Optical bench, feed-through and test stand}\label{sec2}
\begin{figure}[h]
\begin{center}
 \includegraphics[width=7.5cm]{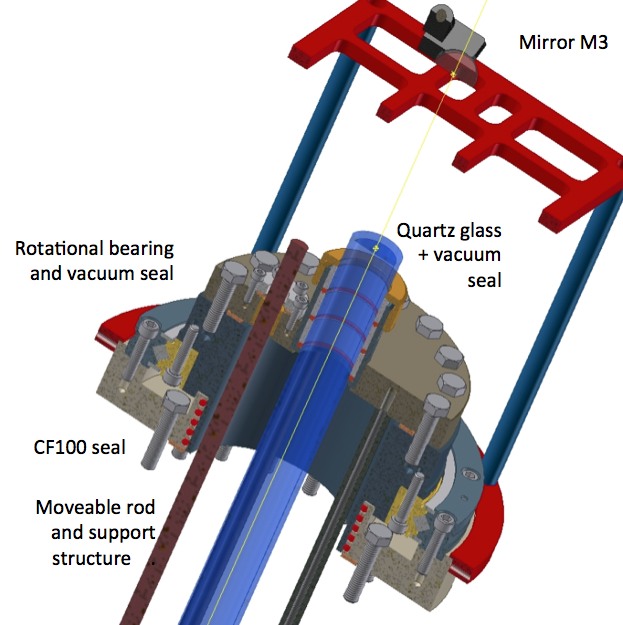}
 \caption{3D CAD sketch of the top part of the feed-through with a sliced view inside the mounting assembly. The glass tube has diameter of 3~cm; the total height of the device is about 1~m.}\label{fig:feedthrough2}
\end{center}
\end{figure}

\begin{figure}[h]
\begin{center}
 \includegraphics[height=5cm]{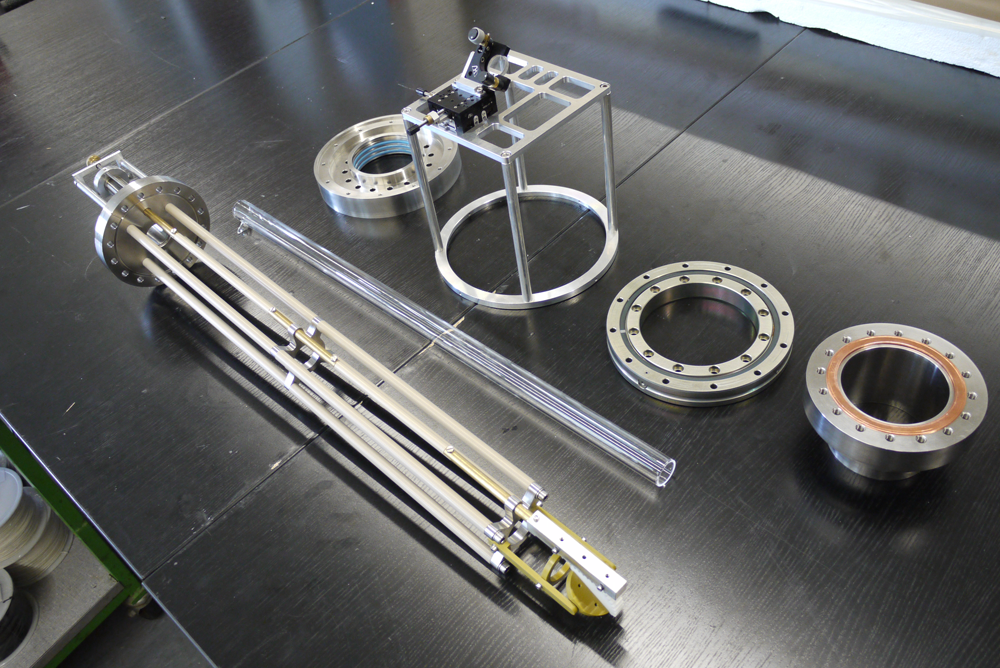}
\scalebox{-1}[+1]{  \includegraphics[height=5cm]{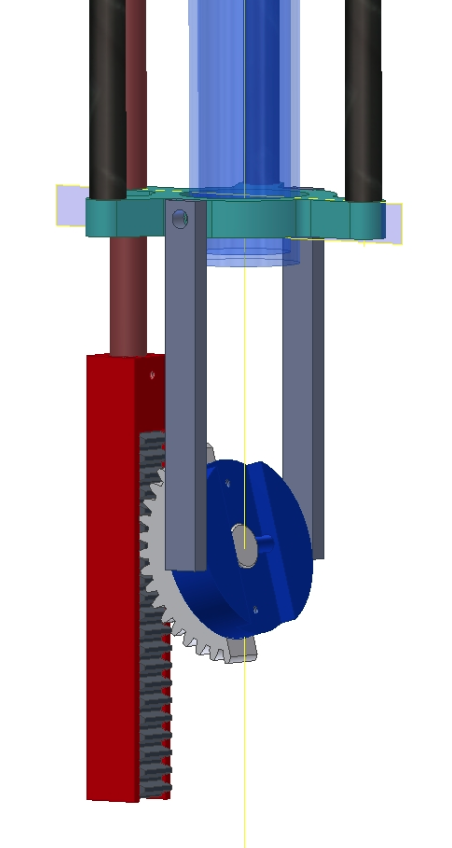} }
 \includegraphics[height=5cm]{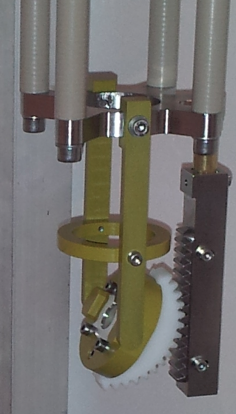}
 \caption{Left: photograph of the feed-through parts. Shown on the bottom is the support structure with the linear rod and the mirror with its cog wheel. Above lies the quartz glass tube. In the top row from left to right are shown the CF100 seals, the top mirror support, the rotational bearing and the connection piece to the top part of the flange. In the middle image a 3D CAD sketch of the bottom part of the feed-through is shown. Right: photograph of the cog wheel, connecting a linear rod and the mirror M4 to vary the angle of the laser beam.}
 \label{fig:feedthrough3}
\end{center}
\end{figure}

The concept of the TPC field mapping is shown in Figure \ref{fig:concept}. An external optical bench allows for the mounting of the UV laser and optics required to provide a path for the laser beam into the liquid argon in the cryostat. A steerable cold mirror (immersed in the cryogenic liquid) allows for different paths into the LAr TPC volume. As a first test we used a prototype TPC with a readout area of 19~cm\,$\times$\,19~cm and an overall drift length of 23~cm. In the case of the MicroBooNE experiment, the LAr TPC will have a much larger size of 2.3~m\,$\times$\,10~m with a maximum drift length of 2.5~m. 

The feed-through needs to be vacuum tight up to 10$^{-4}$~mbar to allow the evacuation of the vessel before filling with liquid argon and the materials used should have an out-gassing in the argon gaseous phase such that the rate of possible impurities degrading the charge lifetime of the liquid argon be lower than the recirculation capabilities of the liquid argon purification system. The feed-through is mounted onto a liquid argon cryostat housing the test TPC. The external optical bench, shown in Figure \ref{fig:opticalbench}, is used to accommodate the Continuum Surelite I-10 UV laser head and optical components to control the laser path. The Surelite I-10 UV laser is emitting light at a wavelength of 1024~nm as the primary light source. Inside the laser head, nonlinear crystals are installed in the beam line for frequency doubling and summing, resulting in a wavelength of 266~nm needed for ionization of liquid argon. For this wavelength, the company specifies an output energy of 60~mJ for each 4 to 6~ns long pulse and a horizontal polarization. The beam spot size at the laser head exit is 5\,mm, the maximal repetition rate is 10~Hz; the beam has a divergence of 0.5~mrad. The Rayleigh scattering for a wavelength of 266\,nm is around 17\,m, the maximum track length is given by the divergence of the laser beam. For alignment purposes, between the optical table and the feed-through mounted on the cryostat, a green light beam from a 532~nm laser diode can enter the beam path at the first fixed dichroic mirror M1. An automated Iris\footnote{Standa Iris 8MID22-0-H} controls the beam spot size from the original size of 5\,mm down to 1\,mm, where deflection became significant. For the measurements presented here we choose a beam spot of 2\,mm as the position resolution for the test TPC was around ~3mm in the X-Y plane (4\,mm wire pitch). An attenuator\footnote{Altechna Watt Pilot Attenuator 2-EWP-R-0266-M} with a turnable $\lambda/2$-plate enables to rotate the orientation of the laser beam polarisation. Behind this two parallel plates are installed such that the angle of the incident beam matches the Brewster Angle of the reflector. Modulating the polarisation of the beam allows to adjust the energy of the reflected beam from 0.3\,\% to 95\,\% of the original beam power. We found that for our test with an incident angle of about 45$^{\circ}$ ~35\,\% of the beam power was required to provide uniform ionization tracks in the test TPC. A remotely operated mirror M2\footnote{Zaber T-OMG, 266~nm dichroic mirror} allows one to finely adjust the beam direction. The laser path length at the optical bench is about 40~cm and the air path between the bench and the feed-through is about 300~cm. We tested the precise steering of the beam over longer distances by guiding the UV laser spot to thermal paper at various distances of up to 12~m. Repeatability turned out to be better than the size of the beam spot. The UV laser box is designed for remote operation and manipulating the beam parameters, further tuning of the beam spot size and intensity will be measured with the data taking of the MicroBooNE experiment.

\begin{figure}[h]
\begin{center} 
 \includegraphics[width=10cm]{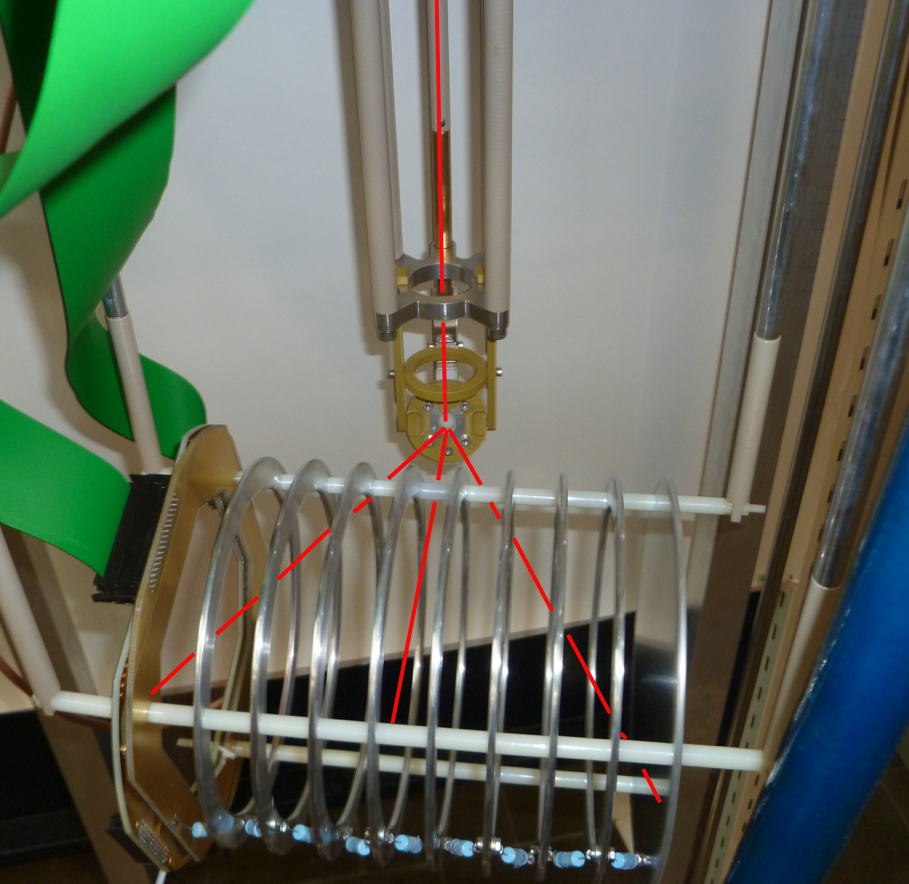}\caption{Photograph of the TPC used for the test. The UV laser feed-through is visible in the back at the centre of the image. Examples of the possible laser beam paths are sketched in red.}\label{fig:tpc}
\end{center}
\end{figure}

The feed-through consists of an evacuated quartz glass pipe with two flat windows, one outside the dewar at room temperature where the laser beam enters, the other reaching well below the liquid surface \cite{uvlaser1}. This allows  the laser beam to enter the liquid argon without being defocused on the uneven liquid/gas argon phase boundary. A sketch of the setup is shown in Figure \ref{fig:feedthrough}. A rotating vacuum seal allows for the movement of a mirror mounted on the bottom of the support structure. The rotational seal was achieved with a custom CF flange. Four silicon rings in the CF100 seal provided vacuum tightness, while a crossed roller bearing provide the structural connection with the vacuum chamber. A linear rod attached to a cog wheel gives the ability to change the angle of the beam entering the TPC. In Figure \ref{fig:feedthrough2} the CAD drawings are shown. Vacuum tightness is realized by four silicon rubber rings fitted into groves along the outer shaft of the top glass holder plate on the feed-through. The whole structure is connected with a rotational bearing to a lower plate, again sealed with rubber silicon rings, for the horizontal rotation for the mirror movement. A CF100 seal connects the feed-through to the cryostat. From the top plate a linear rod extends to the mirror in the liquid argon. A cog wheel allows one to manipulate the mirror M4 and thus the laser beam, as shown in Figure \ref{fig:feedthrough3}. The top part of the rod is welded to a bellow mounted to the top plate on a CF 16 vacuum flange. The rod could move 5\,cm to tilt the mirror 90$^{\circ}$. For the prototype under test the movement was done manually, while in the final design for the MicroBooNE experiment a motorisation is implemented; this will be reported in a forthcoming publication.

The test TPC's sensing wire planes are used to record the drifting charges. sIt has a readout area of 19\,cm$\times$19\,cm and an overall drift length of 23 cm. The wire plane voltage is set such that one is sensing the induction signal of the charges drifting to the readout (induction plane), while the other is set such that the electric field lines end its wires, thus collecting the charges for energy measurement purposes (collecting plane). Every other readout wire is instrumented with pre-amplifiers placed outside the signal feed-through. Field shaping rings are connected via a resistor chain to a high voltage feed-through. The shaping rings create sufficiently wide gaps for the UV laser beam entering the TPC from the side, as shown in Figure \ref{fig:tpc}, and generate tracks analogous to those produced \textit{e.g.} by cosmic ray muons. The DAQ system consists of CAEN V1729 digitisers. As trigger for the DAQ, the signal of a photodiode\footnote{Thorlabs Det10A} exposed to the UV laser beam was used, as sketched in Figure \ref{fig:opticalbench}. The cryostat and DAQ are the same as in \cite{uvlaser1,uvlaser2}. Given the rate of UV laser generated tracks, background from cosmic ray muons, target mass and purity are irrelevant for this specific performance test.

\section{Performance of the prototype}\label{sec3}

\begin{figure}[h]
\begin{center} 
\includegraphics[width=11cm]{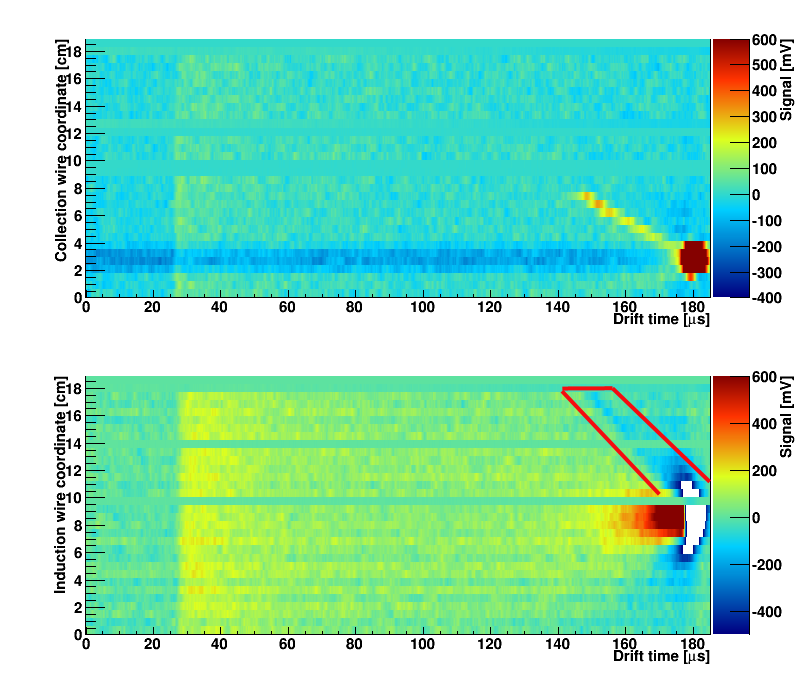}
\end{center}
\caption{Image of the UV laser run. On the abscissa the drift time is given, on the ordinate the wire coordinate in cm. The colour scale indicates the charge recorded on each wire. The laser enters the chamber such that in the collection plane view only wires 0 to 8 are sensitive; for the collection plane view the laser induces signals on wire 10-18 before hitting the cathode. This was due to the mirror inclination used to obtain these images of a horizontal sweep. A red box indicated the signal induced by the UV laser track in the induction view. Note the deposited charge due to the photoelectric effect from the UV laser hitting on the cathode.}
\label{fig:data1}
\end{figure}

\begin{figure}[t]
\begin{center}
\includegraphics[width=11cm]{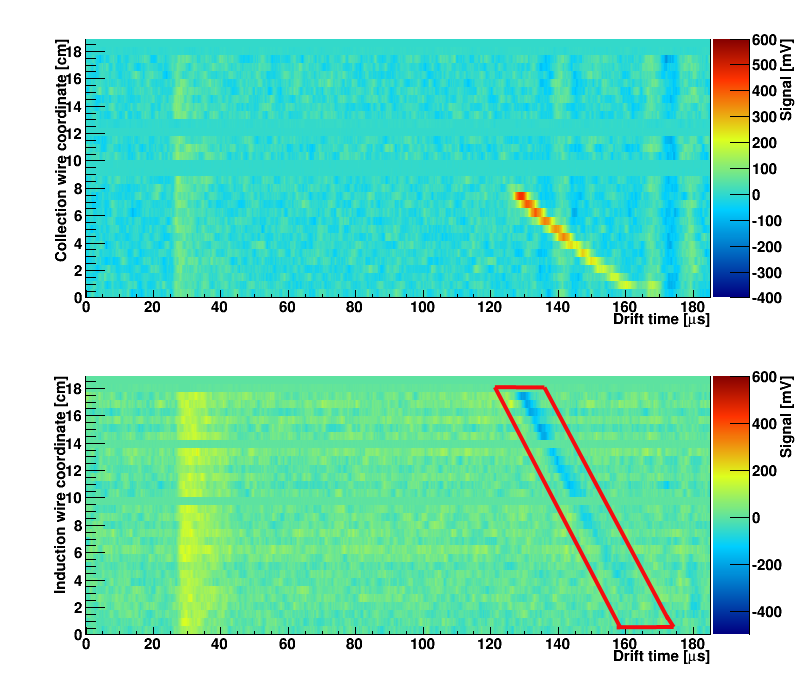}
\end{center}
\caption{Image of the UV laser run. On the abscissa the drift time is given, on the ordinate the wire coordinate in cm. The colour scale indicates the charge recorded on each wire. The laser enters the chamber such that in the collection plane view only wires 0 to 8 are sensitive, while for the induction wire plane with perpendicular orientation the full 18 wires are exposed. A red box indicated the signal induced by the UV laser track. This images shows the laser track after rotating the feed-through by about 15~degrees.}
\label{fig:data2}
\end{figure}

All components used for the mirror support structure located in the cryostat were screened at liquid argon temperature and in the gas phase with a test stand at FNAL. Their influence on the pollution for the liquid argon is at a negligible level. The feed-through was tested for vacuum tightness and reached $10^{-4}$~mbar, also while being rotated. The measured leakage rate was less than $10^{-8}$~mbar$\cdot$l/s as measured with a helium leak tester. 

The argon used to fill the TPC was purified using standard Oxisorb and Hydrosorb cartridges. The charge lifetime reached in this measurement is about 0.2~ms, fairly adequate for our scopes. As for this test no standard recirculation and further purification of the argon was done, the purity decreased over time, such that after a period of 2 days the charge lifetime dropped such that no signal from the cathode area could be reconstructed anymore. This is in good agreement with previous tests performed with this cryostat \cite{uvlaser1,uvlaser2}.

The mirrors of the optical bench and the feed-through were positioned and aligned before cooling down to confirm that a free path into the TPC existed. After cooling down, a re-adjustment was required due to the thermal shrinkage of the support structure by using the alignment laser. The main alignment that needed to be corrected was between feed-through and optical bench, the only adjustment for the optical path into the test TPC was a small rotation to clear the obstruction from a field cage ring.

Data taking with the UV laser was performed for different cold mirror configurations by varying the inclination angle, as well as rotating the assembly to cover the whole test TPC volume. Two typical events are shown in Figures \ref{fig:data1} and \ref{fig:data2}. The laser beam enters from the center of the TPC such that in the view of the collected charge only wires 0 to 8 are sensitive, while for the induction wire plane with perpendicular orientation the full 18 wires are exposed. This was determined by the actual mirror configuration when acquiring these events. The UV laser hitting the cathode is also visible from the detection of the 'splash' of charge produced by photoelectric effect on the stainless steel. This charge clouds allows to extract precisely the distance between wire readout and cathode in the units of the drift time. The UV laser intensity was tuned such that the event is visible by eye in the collection view for the images presented here, It was not optimized for the calibration measurement, neither recombination, nor purity corrections were applied. As the positioning of this device was done by hand, no measurement of pointing accuracy was obtained better than the 0.5\,mm resolution from a meter attached to the feed-through. We studied the data on an event-by-event base to tune the beam intensity for visibility by eye, no other performance parameters were extracted for this setup. In a further publication we will present the results of the MicroBooNE feed-through, where we upgraded the system using commercial CF components for the rotational seal and linear actuator from Thermionics, USA. This two manipulators are certified for the use in UHV application, due to the mounting of the rotational seal with a worm drive the backslash on the rotational movement can be suppressed. The reason for this upgrade is to allow remote operation of the system, the certified leak tightness, as well as the improved positioning resolution provided with an automated optical readout system by Haidenhein with sub-micron resolution. Further studies on the ionization density and laser beam parameters are foreseen during the commissioning of the MicroBooNE experiment.

\section{Conclusion}
We presented here the design and the prototype tests of the first system able to steer a UV laser beam into a LAr TPC. The design allows a full 3D movement of the beam and can be well applied to larger size LAr TPC's as a field mapping calibration system. Straight ionizing tracks are generated by an UV laser beam, apart from regions inaccessible due to presence of the field shaping rings of the TPC cage. The assembled device is a prototype of the system that we are implementing for the MicroBooNE experiment at FNAL. UV laser generated tracks can also be used to measure liquid argon properties such as the purity \cite{uvlaser2} or to perform charge diffusion measurements. The latter will be the subject of a forthcoming publication.

\section{Acknowledgements}
We warmly thank S. Pordes and E. Skup from FNAL for their kind support in the out-gassing measurements of the detector components. We also thank our engineer R. H\"{a}nni for the design and the LHEP workshop for the construction of the prototype. We thank the Bern EXO group for sharing TPC parts with us.

\end{document}